# GIAO-DFT Isotropic magnetic shielding constants and spin-spin coupling of tartaric acid in water solution

**Bruna Fideles, Leonardo B. A. Oliveira and Guilherme Colherinhas***

Departamento de Física - CEPAE, Universidade Federal de Goiás, CP 131, 74001-970, Goiânia, GO, Brazil.

[*] Corresponding author. E-mail: gcolherinhas@gmail.com, Fax: +55.62.3521-1083

# GRAPHICAL ABSTRACT

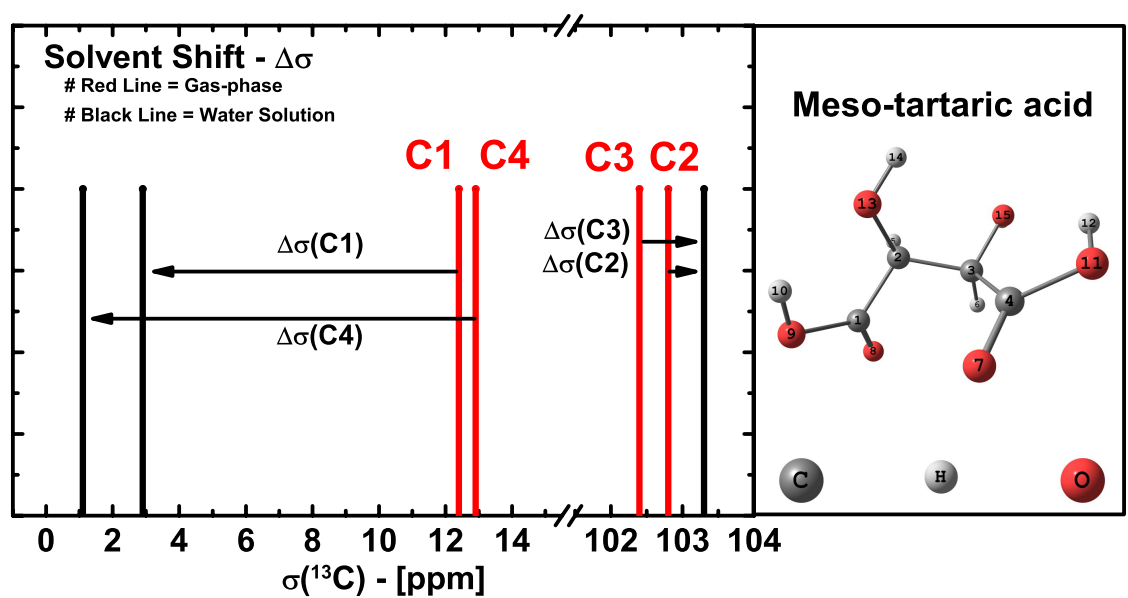

## ABSTRACT

We investigate the nuclear isotropic shielding constants and spin-spin coupling for oxygen and carbons atoms of isomers of tartaric acid in gas phase and water solutions by Monte Carlo simulation and quantum mechanics calculations using the GIAO-B3LYP approach. Solute polarization effects are included iteratively and play an important role in the quantitative determination of shielding constants. Our MP2/aug-cc-pVTZ results show substantial increases of the dipole moment in solution as compared with the gas phase results (61-221%). The solvent effects on the $\sigma(^{13}C)$ [J(C-C)] values are in general small. More appreciable solvent effects can be seen on the $\sigma(^{17}O)$ and J(C-O).

**HIGHLIGHTS:** GIAO-DFT-NMR, Solvent effects, Sequential MC/QM, shielding constants, spin-spin coupling constants.

# 1. INTRODUCTION

In general, optical isomerism is a type of spatial isomerism that can be observed when there are chiral carbons in the molecular structure, this fact introducing an asymmetry in the molecule responsible for isomerism. The isomers with this type of structure are more commonly known as enantiomers. Among the various enantiomers, we highlight the meso tartaric acid isomers (TA). Because of the effect of two chiral carbon atoms in its molecular structure, tartaric acid has three different optical isomers: dextro-tartaric acid (DTA), levo-tartaric acid (LTA) and meso-tartaric acid (MTA) (see Figure 1). For example, LTA are usually found in wine, responsible for determining its acidity and used as a regulator of its concentration of pH; LTA or DTA are the requirements in food and drug quality and management [01]. The difference of these isomers is apparently subtle but provide influences on geometrical, electrical and magnetic properties.

In recent work, Wenge Yang et al studied the solubility of LTA in pure ethanol, propanol, isopropanol, n-butanol, acetone and acetonitrile in the temperature range 281.15 and 324.25 K [02]. Ralf Tonner et al investigated the stability of gas phase TA anions using quantum chemistry, mass spectrometry and infrared spectroscopy; the study presented a comprehensive analysis of the thermodynamic and kinetic stability of the tartaric acid dianion [03]. A detailed $^1$H NMR analysis of vinegars (including tartaric acid) is reported by A. Caligiani et al. The study described can be satisfactorily used for the quantification of several organic compounds in vinegar, in particular organic acids [04]. Z. Dega-Szafran et al investigated recently a molecular structure of the hydrated complex of trigonelline with tartaric acid. The authors report the NMR spectra of the elucidated structure of the complex in aqueous solutions [05].

Following the line of magnetic properties studies, the magnetic shielding constant and spin-spin coupling are sensitive properties that can distinguish between isomers [06,07]. These properties are fundamentally influenced by the presence of the solvent and also by the geometrical structure of the compound in solution. From a theoretical point of view the inclusion of solvent effects on the geometry of the solute can be made using the polarizable continuum model (PCM). However, the uses of continuous models for describing solute-solvent interactions are quite poor. In general, the geometry obtained with PCM is considered as a starting point for computer simulations in order to describe the solvent discretely, for example, simulations using the classical Monte Carlo [08-10].

In this work, the isotropic magnetic shielding and spin-spin coupling constants for carbon and oxygen atoms of the dextro-tartaric acid (DTA), levo-tartaric acid (LTA) and meso-tartaric acid (MTA) – see Figure 1 – in water are based on the sequential Monte Carlo simulation/quantum mechanics (S-MC/QM) methodology [08-10]. In this approach, the MC simulation generates solute-solvent configurations and quantum mechanical calculations are performed on these configurations composed of one solute molecule surrounded by several solvent molecules treated as point charges or/and as explicit molecules. For the NMR properties in gas-phase and water, we use the density functional theory (DFT) and gauge invariant atomic orbital (GIAO) [11,12] method - GIAO-DFT approach - which offers a good compromise between computational cost and accuracy [13,14]. We verify that all values reported here are statistically converged.

We develop the solute polarization effect for describing the NMR constants of isomers of tartaric acid in water solvent. For NMR calculations we consider configurations with explicit water molecules hydrogen bonded (HB) with the solute. Thus, one realistic contribution of the solvent effects is used to obtain the magnetic properties of the solute.

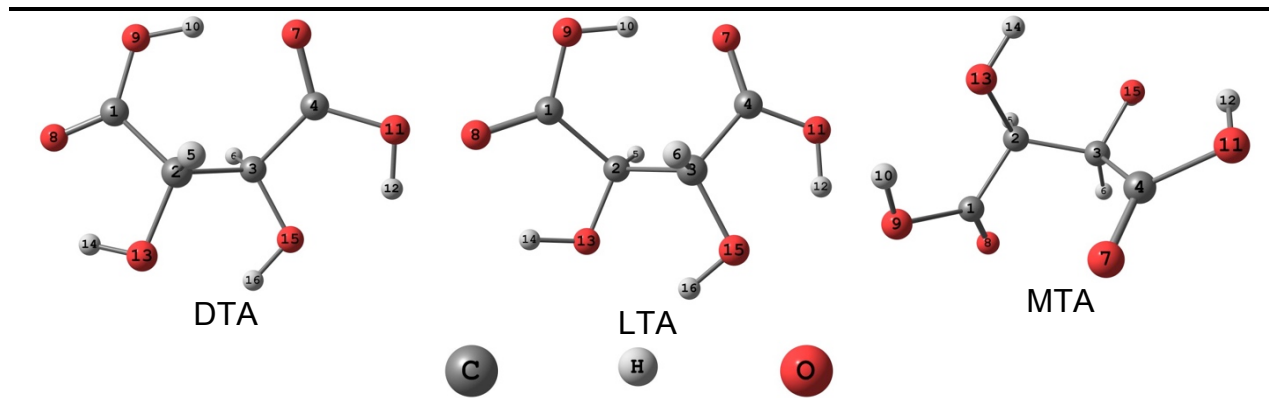

**Figure 1:** Dextro-tartaric acid (DTA), levo-tartaric acid (LTA) and meso-tartaric acid (MTA) geometry obtained with B3LYP/cc-pVTZ.

## 2. COMPUTATIONAL DETAILS

The ground state geometry of isomers of tartaric acid in gas-phase and in-water were fully optimized with the B3LYP exchange–correlation functional with the cc-pVTZ basis set, without any symmetry constraint. Initially, the solvent dependence of the geometry has been included by employing the self-consistent reaction field (SCRF) approach with the polarizable continuum model (PCM) [15-17] as implemented in the GAUSSIAN 09 package [18].

The Monte Carlo (MC) simulations were performed for all molecules through DICE program [19] by using the Metropolis sampling technique in the *NPT* ensemble for a system composed by one molecule of tartaric acid and 1000 molecules of water in temperature of 300.15 K and pressure of 1 atm. Solute and solvent molecules are held with rigid geometry during the MC simulations but the solute geometry used in the MC simulations during whole polarization process was optimized in water using SCRF-PCM model. In the MC simulation the intermolecular interactions are modeled by the standard Lennard-Jones (LJ) plus Coulomb potential with three parameters for each atomic site ($\varepsilon_i$, $\sigma_i$, and $q_i$) and the combination rules are $\varepsilon_{ij} = (\varepsilon_i \varepsilon_j)^{1/2}$ and $\sigma_{ij} = (\sigma_i \sigma_j)^{1/2}$. For the solute, we have used the LJ parameters of the optimized parameters for liquid simulation (OPLS) force field [20] (displayed in Table 1) and the MP2 (second-order Møller–Plesset perturbation theory), with the aug-cc-pVTZ basis set. Atomic charges were obtained by using an electrostatic potential fit (CHELPG) [21]. For water the simple point charge (TIP3) model from ref. [22] was used as force field.

We start the iterative S-MC/MQ procedure (SIM-1) by performing a MC simulation with the coulombic term of the solute potential described by its CHELPG atomic charges obtained in water using SCRF-PCM model at the MP2/aug-cc-pVTZ level. To initiate another step of the iterative process, we select statistically uncorrelated configurations to generate an electrostatic embedding around the solute. The average atomic charges of the solute are calculated using the average solvent electrostatic configuration (ASEC) [23]. In all cases, the ASEC is composed by superposition of 100 statistically uncorrelated configurations bearing a solute molecule and 400 solvent molecules, treated as point charges. This procedure is repeated until convergence of the solute dipole moment in solution is achieved. The results obtained with these configurations are referred to as polarized (POL) model results. In general, the dipole moment values of the solute in solution present a rapid convergence pattern as function of the iterative step [24-26].

**Table 1:** Lennard-Jones potential parameters of the Monte Carlo simulation. $\varepsilon$ (kcal mol$^{-1}$) and $\sigma$ (Å).

| Atoms  | $\varepsilon$ | $\sigma$ |
|--------|--------|--------|
| C1, C4 | 0.1050 | 3.7500 |
| C2, C3 | 0.0660 | 3.5000 |
| H (–C) | 0.0300 | 2.5000 |
| H (–O) | 0.0000 | 0.0000 |
| O (=C) | 0.2100 | 2.9600 |
| O (–H) | 0.1700 | 3.0000 |

In additional, after the polarization process, we selected 100 uncorrelated configurations with some water molecules that form HBs with solute, considering the following criteria: maxima distance between the sites ($O_{Solute} \ldots O_{Solvent}$) less than or equal to 3.3 Å, with $\angle O \ldots O - H \leq 40°$ and the solute-solvent interaction energy less than –2.5 kcal/mol, and more 400 solvent molecules treated as point charges, for each isomer. The results obtained with these configurations are referred to as polarized/hydrogen bonds/charge points (POL*) model results.

The $\sigma$ (isotropic magnetic shielding, in ppm) and $J$ (spin-spin coupling, in Hz) constants for carbon and oxygen atoms were calculated for gas phase, POL and POL* models using the GIAO approach using the B3LYP exchange–correlation functional with the 6-311++G(2d,2p) basis set, as implemented in the GAUSSIAN 09 program [18].

## 3. RESULTS AND DISCUSSION

### 3.1 Optimized geometric parameters.

The B3LYP/cc-pVTZ bond lengths (R) for isomers of tartaric acid (DTA, LTA and MTA) optimized in gas-phase and in water solution are presented in Table 2. We can estimate the solvent effect on the geometries of isomers of TA observing the results of $\Delta$ (= $R_{Gas\ phase} - R_{Water\ PCM}$). For DTA and LTA the results are equal, in this case the more sensitive bond lengths are R(C1-O9), R(C2-C3), R(C2-O13), R(C4-O11), R(O9-H10) and R(O13-H14) with results of $\Delta$ equal to -0.022, 0.013, -0.012, -0.010, 0.027 and 0.011 Å, respectively. For these molecules, the other values of R show results of $\Delta$ between -0.001 and 0.007 Å. In contrast, the results obtained for the MTA molecule are between -0.005 and 0.003 Å, only the lengths R(C3-O15), R(C4-O7) and R(C4-O11) have different values, equal to 0.012, -0.009 and 0.008 Å respectively, being more sensitive to solvent effects.

**Table 2:** B3LYP results for selected bond distances (in Å) of the isomers of tartaric acid (TA) optimized in gas-phase and in-water solution (with PCM/cc-pVTZ). $\Delta = R_{Gas\,phase} - R_{Water\,PCM}$.

| R | DTA | | | LTA | | | MTA | | |
|---|---|---|---|---|---|---|---|---|---|
|  | Gas phase | Water | Δ | Gas phase | Water | Δ | Gas phase | Water | Δ |
| R(C1-C2) | 1,540 | 1,540 | 0,000 | 1,540 | 1,540 | 0,000 | 1,533 | 1,531 | 0,002 |
| R(C1-O8) | 1,208 | 1,202 | 0,006 | 1,208 | 1,202 | 0,006 | 1,200 | 1,205 | -0,005 |
| R(C1-O9) | 1,320 | 1,342 | -0,022 | 1,320 | 1,342 | -0,022 | 1,333 | 1,333 | 0,000 |
| R(C2-C3) | 1,557 | 1,544 | 0,013 | 1,557 | 1,544 | 0,013 | 1,532 | 1,535 | -0,003 |
| R(C2-H5) | 1,095 | 1,093 | 0,002 | 1,095 | 1,093 | 0,002 | 1,095 | 1,092 | 0,003 |
| R(C2-O13) | 1,406 | 1,418 | -0,012 | 1,406 | 1,418 | -0,012 | 1,418 | 1,421 | -0,003 |
| R(C3-C4) | 1,524 | 1,525 | -0,001 | 1,524 | 1,525 | -0,001 | 1,536 | 1,533 | 0,003 |
| R(C3-H6) | 1,097 | 1,094 | 0,003 | 1,097 | 1,094 | 0,003 | 1,092 | 1,091 | 0,001 |
| R(C3-O15) | 1,412 | 1,413 | -0,001 | 1,412 | 1,413 | -0,001 | 1,435 | 1,423 | 0,012 |
| R(C4-O7) | 1,213 | 1,206 | 0,007 | 1,213 | 1,206 | 0,007 | 1,197 | 1,206 | -0,009 |
| R(C4-O11) | 1,322 | 1,332 | -0,010 | 1,322 | 1,332 | -0,010 | 1,340 | 1,332 | 0,008 |
| R(O9-H10) | 0,993 | 0,966 | 0,027 | 0,993 | 0,966 | 0,027 | 0,971 | 0,976 | -0,005 |
| R(O11-H12) | 0,976 | 0,976 | 0,000 | 0,976 | 0,976 | 0,000 | 0,972 | 0,977 | -0,005 |
| R(O13-H14) | 0,975 | 0,964 | 0,011 | 0,975 | 0,964 | 0,011 | 0,963 | 0,962 | 0,001 |
| R(O15-H16) | 0,970 | 0,968 | 0,002 | 0,970 | 0,968 | 0,002 | 0,962 | 0,963 | -0,001 |

**3.2 Solute polarization**.

**Figure 2:** MP2/aug-cc-pVTZ results for the dipole moment (in D) of isomers of tartaric acid (TA) in gas-phase (GAS) and in-water solution – PCM model and Polarized model. Evolution of the results for $\mu$ of all isomers of TA as function of the number of S-MC/QM iterations The dipole moments of solution are calculated with PCM and S-MC/QM process using ASEC solvation model (POL).

| Isomers | GAS | PCM | POL |
|---|---|---|---|
| MTA | 6.77 D | 9.38 D | 10.90 D |
| DTA | 3.12 D | 8.27 D | 9.93 D |
| LTA | 3.12 D | 8.27 D | 10.00 D |

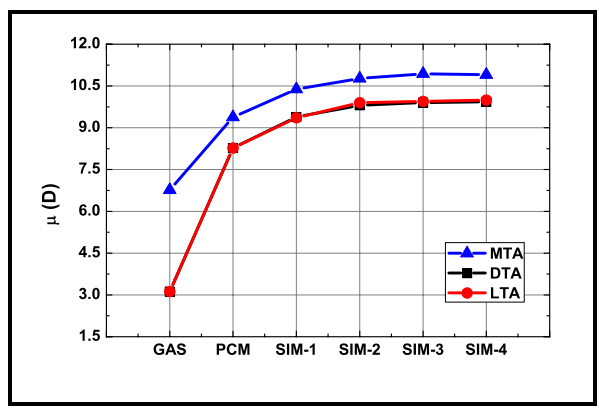

MP2/Aug-cc-pVTZ results for the dipole moment ($\mu$) obtained in gas-phase (GAS) and in-water solution with the PCM and S-MC/QM polarized model (POL), are quoted in Figure 2. In the Figure 2 we show the evolution of the results for $\mu$ of all isomers of TA as function of the number of S-MC/QM iterations. Each displayed point (SIM-#) corresponds to a statistically converged result obtained from 100 uncorrelated configurations ($< 10\%$) using the ASEC solvation model where the solvent molecules are treated as point charges. One can see that the convergence of the dipole moment with respect to the number of iterations is rapid, as shown for other molecular systems in solution [24-26]. The PCM dipole moment is showed for comparison.

For the DTA, LTA and MTA molecules the gas phase MP2/Aug-cc-pVTZ values of $\mu$ are equal to 3.12, 3.12 and 6.77 D, respectively. In solution, these results obtained using POL [PCM] model are equal to 9.93, 10.00 and 10.90 D [8.27, 8.27 and 9.38 D] respectively. The comparisons between gas phase and solution results showed increases in the values of $\mu$ equal to 218%, 221% and 61% [165%, 165% and 39%].

### 3.3 Hydrogen bonds.

Previous studies have shown that the polarization effect of the solute can affect considerably the specific interactions between solute and solvent molecules [27]. Figure 3 shows radial distribution functions (RDF) between the oxygen atoms of the TA isomers and the oxygen atom of the water molecule obtained with the POL solute models. The first peaks of the RDF are centered in typical length of HBs between solute and solvent (2.5 - 3.0 Å). The system presents considerable differences for the RDFs, e.g. $G_{O9-O}(r)$ shows that the first coordination shell for DAT and LAT isomers is narrower than in the MAT isomer. This occurs due to isomeric changes of the compounds that modify form the first layer of solvation. The integration of the first coordination shell, we obtain the average numbers of water molecules closest to the oxygen atoms of the TA isomers. For example, the isomers DTA [MTA] show 2.09, 2.75, 1.42, 4.26, 4.09 and 4.47 [2.54, 2.07, 6.33, 7.22, 1.19 and 1.26] water molecules close to the atoms O7, O8, O9, O11, O13 and O15, respectively. Due to proximity between the oxygen atoms of isomers, these results may have overlapping in the counting of water molecules. Thus, we also present a quantitative analysis of the average number of HBs in aqueous solution using geometric and energetic criteria. In this case, we use 100 uncorrelated configurations for selection of water molecules that form HBs. For the DTA [MTA] molecule the average number of HBs between O7, O8, O9, O11, O13 and O15 and water is of 1.52, 1.82, 1.50, 0.39, 1.33 and 0.66 [1.78, 1.70, 0.46, 0.47, 1.02 and 1.02] by setting, respectively. The data for the LTA isomer are similar to DTA. Figure 4 illustrates for the TA isomers the configurational space of HBs in aqueous solution, obtained from 100 uncorrelated configurations. The treatment of explicit molecules can be particularly relevant for a reliable determination of the shielding constants.

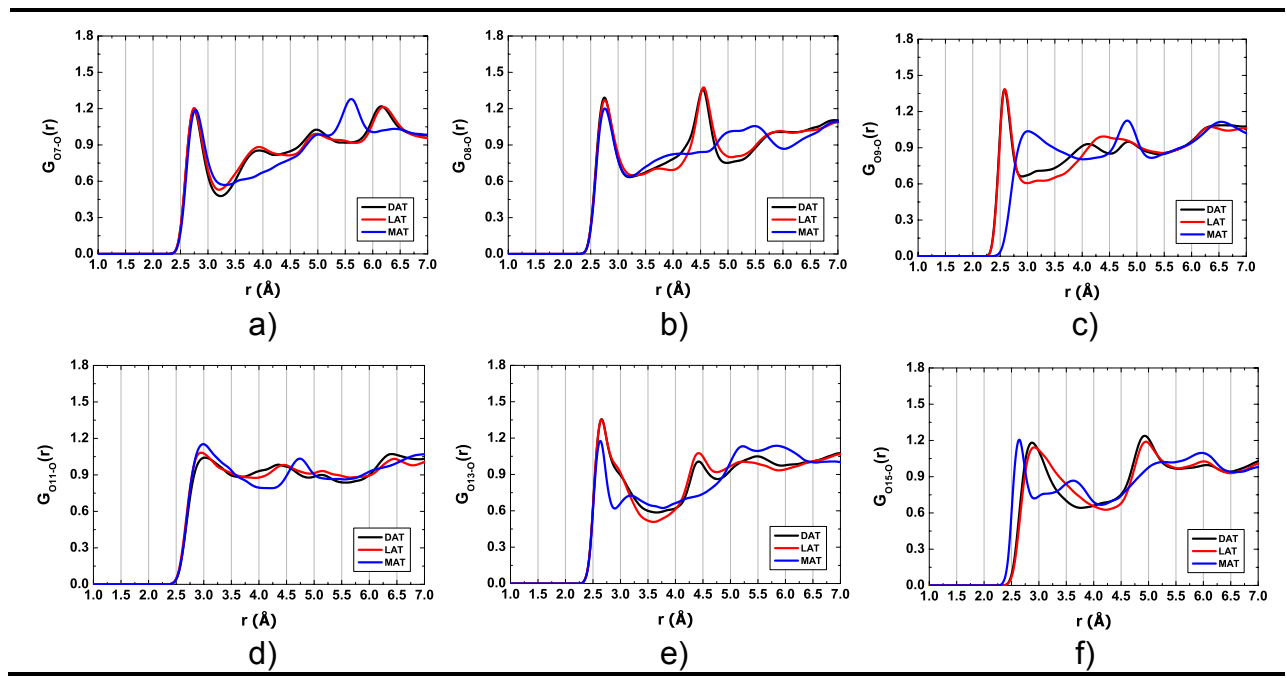

**Figure 3:** Radial distribution function ($G_{OX-O}(r)$) for the a) O7; b) O8; c) O9; d) O11; e) O13 and f) O15 oxygen atoms of tartaric acid isomers. Red line (LAT), blue line (MAT) and black line (DAT).

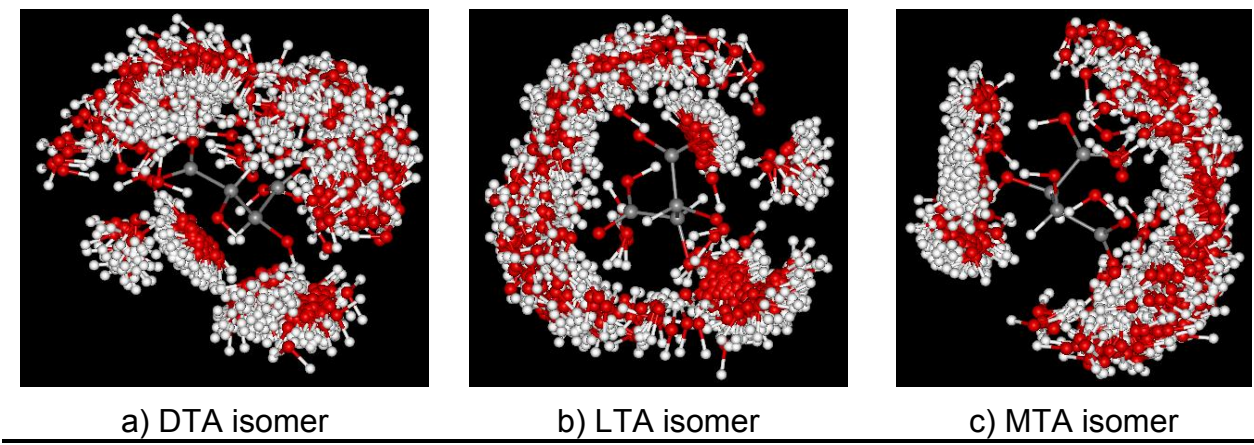

| a) DTA isomer | b) LTA isomer | c) MTA isomer |

**Figure 4:** Configurational HBs space for the TA isomers the in aqueous solution obtained from 100 uncorrelated configurations.

### 3.4 Shielding constants in gas-phase and in-solution.

Next sections shows the solvent effects on $\sigma(^{13}C)$ and $\sigma(^{17}O)$ of the TA isomers. The shielding constants in solution were obtained using the ASEC solvation model. PCM results are presented for comparison.

### 3.4.1 Solvent effects on $\sigma(^{13}C)$

Table 3 shows the results for gas phase and in-water polarized shielding constants obtained at the GIAO-B3LYP/6-311++G(2d,2p) level for the TA isomers. An estimation of the solvent shift for the $\sigma$ can be analyzed from the differences between the results obtained in-solution and gas phase ($\Delta\sigma = \sigma_{SOLVENT} - \sigma_{GAS}$).

The results shows that the presence solvent does not affect significantly the shielding constant values for carbons C3 and C4. The $\Delta\sigma_{POL}$ is close to -3 and 2 [-3 and 2] ppm, unlike presented, the results for C1 and C2 shows little influence of the solvent: results for $\Delta\sigma_{POL}$ is close to -6 [-6] ppm for DTA [LTA]. For MTA there is an inversion, for C1 and C4 the $\Delta\sigma_{POL}$ is close to -10 and -12 ppm e for C2 and C3 a insignificant influence of the 1 ppm. This is observed because the form of exposure of the carbon atoms of the middle isomers solvent apparently in the MTA configuration there is a greater exposure to the solvente medium of the terminals carbon atoms, which constitute the carbon skeleton of the molecule.

To examine the influence of HBs on $\sigma$ in aqueous solution, we consider 100 uncorrelated supermolecular structures obtained from the MC simulations with the explicit inclusion of some hydrogen-bonded water molecules embedded in a solvation shell composed by 400 outer water molecules, treated as point charges (Tip3 model). The results obtained with these configurations are referred to as POL* model in Table 4. The results of Table 4 shows that the presence of explicit water molecules does not affect the $\sigma$ values. The $\Delta\sigma_{POL*}$ values for C1, C2, C3 and C4 are of -6, -4, -3 and 2 ppm for DTA, -6, -4, -3 and 1 ppm for LTA and -10, 1, 2 and -13 for MTA isomer.

**Table 3:** Gas phase and in-water polarized shielding constants [ppm] and Spin-Spin coupling contants [Hz] obtained at the GIAO-B3LYP/6-311++G(2d,2p) for carbon and oxygen atoms of the tartaric acid isomers. GAS (gas phase model), POL (polarized/point charge model), POL* (polarized/hydrogen bonds/point charge model), PCM (polarized continuum model).

| Atom | Magnetic Shielding Constants [ppm] | | | | | | | | | | | |
|---|---|---|---|---|---|---|---|---|---|---|---|---|
| | DTA | | | | LTA | | | | MTA | | | |
| | GAS | POL | POL* | PCM | GAS | POL | POL* | PCM | GAS | POL | POL* | PCM |
| C1 | 6.1 | 0.0 | -0.1 | 3.9 | 6.1 | 0.2 | 0.1 | 3.9 | 12.4 | 2.9 | 2.5 | 5.6 |
| C2 | 111.0 | 105.3 | 107.0 | 104.6 | 111.0 | 105.3 | 107.0 | 104.6 | 102.8 | 103.3 | 103.9 | 103.3 |
| C3 | 108.7 | 106.1 | 105.9 | 106.4 | 108.7 | 106.0 | 105.7 | 106.4 | 102.4 | 103.3 | 104.1 | 103.5 |
| C4 | 0.0 | 2.4 | 1.9 | 5.2 | 0.0 | 1.6 | 1.1 | 5.2 | 12.9 | 1.1 | 0.4 | 4.9 |
| O7 | -50.4 | -36.2 | -40.6 | -60.7 | -50.4 | -31.2 | -37.4 | -60.7 | -91.6 | -27.2 | -32.6 | -55.1 |
| O8 | -67.1 | -40.9 | -48.7 | -72.2 | -67.1 | -45.1 | -52.1 | -72.2 | -79.3 | -34.7 | -41.7 | -53.8 |
| O9 | 108.4 | 102.5 | 80.7 | 106.1 | 108.4 | 105.2 | 84.2 | 106.1 | 127.8 | 121.4 | 115.8 | 124.5 |
| O11 | 119.4 | 114.8 | 109.8 | 124.4 | 119.4 | 115.5 | 109.9 | 124.4 | 123.1 | 111.3 | 105.6 | 118.1 |
| O13 | 286.9 | 281.7 | 263.0 | 273.9 | 286.9 | 281.1 | 262.1 | 273.9 | 293.0 | 296.3 | 277.3 | 290.7 |
| O15 | 288.3 | 288.7 | 280.7 | 287.0 | 288.3 | 288.3 | 279.0 | 287.0 | 279.1 | 286.9 | 269.4 | 281.4 |

| $J$ | SPIN-SPIN Coupling Constants [Hz] | | | | | | | | |
|---|---|---|---|---|---|---|---|---|---|
| | DTA | | | LTA | | | MTA | | |
| | GAS | POL | PCM | GAS | POL | PCM | GAS | POL | PCM |
| $C1-C2$ | 47.6 | 57.4 | 58.2 | 47.6 | 57.5 | 58.2 | 58.4 | 56.3 | 57.0 |
| $C2-C3$ | 34.2 | 36.9 | 37.2 | 34.2 | 36.9 | 37.2 | 41.7 | 43.8 | 44.0 |
| $C3-C4$ | 54.2 | 57.6 | 57.9 | 54.2 | 57.4 | 57.9 | 57.6 | 56.8 | 56.8 |
| $C1-O8$ | 26.0 | 21.4 | 24.4 | 26.0 | 21.9 | 24.4 | 24.6 | 22.6 | 24.7 |
| $C1-O9$ | 25.7 | 29.1 | 29.5 | 25.7 | 29.0 | 29.5 | 29.0 | 28.4 | 28.7 |
| $C2-O13$ | 19.6 | 17.7 | 18.2 | 19.6 | 17.7 | 18.2 | 20.4 | 18.3 | 18.8 |
| $C3-O15$ | 19.8 | 20.1 | 20.5 | 19.8 | 20.2 | 20.5 | 19.5 | 18.4 | 18.7 |
| $C4-O7$  | 24.6 | 22.5 | 25.0 | 24.6 | 22.1 | 25.0 | 25.0 | 22.5 | 25.1 |
| $C4-O11$ | 28.1 | 29.4 | 29.2 | 28.1 | 29.2 | 29.2 | 30.3 | 28.8 | 29.1 |

### 3.4.2 Solvent effects on $\sigma(^{17}O)$

In Table 3 we show gas-phase and in-solution results for $\sigma(^{17}O)$ computed with the GIAO-B3LYP/6-311++G(2d,2p) method for TA isomers. The results shows that $\sigma(^{17}O)$ values presents a solvent dependence. We observe that for O7 and O8 there is a larger solvent shift on $\sigma(^{17}O)$ values calculated when POL models, while for O9, O11, O13 and O15 we have a low solvent shift for $\sigma(^{17}O)$. The results for $\Delta\sigma_{POL}$ are close to 14, 26, -6 -5 -5 and 0 [19, 22, -3, -4, -6 and 0] ppm for DTA [LTA] isomers, and more intensively to MTA isomer, whose values are next to 64, 45, -6 -12, 3 and 8 ppm, respectively to O7, O8, O9, O11, O13 and O15. This behavior shows the difference between O7 and O8 atoms from the others showing a typical magnetic signature for the carbonyl [hidroxyl] atoms type, where $\sigma(^{17}O)$ values increase [decrease] with the addition of solvent, except for the O13 and O15 of the MTA isomer.

As expected, the influence of HBs on $\sigma(^{17}O)$ is great, providing a redistribution of magnetic shielding of the oxygen atoms. For the results obtained from the POL* model we observed a inversion in the intensity of the $\Delta\sigma_{POL*}$ values for the oxygen atoms of the isomers DTA and LTA. In these case, the $\Delta\sigma_{POL*}$ values for DTA are next to 10, 18, -28, -10, -24 and -8 ppm, and for LTA are next to 13, 15, -24, -10, -25 and -9 ppm, respectively to O7, O8, O9, O11, O13 and O15. The results for MTA isomes prevail the same trends observed earlier, $\Delta\sigma_{POL*}$ values are next to 59, 38, -12, -18, -16 and -10 ppm. Although changed, the results obtained with POL and POL* models show the same trends with increases [reductions] in the values of $\sigma(^{17}O)$ for the =O [-OH] atom type. We emphasize that the results with PCM model shows different trends for $\sigma(^{17}O)$ values of the DTA and LTA isomers.

### 3.5 SPIN-SPIN coupling in gas-phase and in solution.

Next sections shows the solvent effects on $J(C-C)$ and on $J(C-O)$ of the TA isomers. The SPIN-SPIN coupling constants in solution were obtained using the ASEC solvation model. PCM results are presented for comparison.

### 3.5.1 Solvent effects on $J(C-C)$.

Table 3 shows the results for gas phase and in-water polarized SPIN-SPIN coupling constants obtained at the GIAO-B3LYP/6-311++G(2d,2p) level for the TA isomers. An estimation of the solvent shift for the $J$ can be analyzed from the differences between the results obtained in solution and in gas phase ($\Delta J = J_{SOLVENT} - J_{GAS}$).

Overall the results for $J(C-C)$ is greater than $J(C-O)$ in gas phase and in-water solution. For $J(C-C)$ of the isomers DTA and LTA we obtained similar results with each other. The differences between solution and gas phase results ($\Delta J_{POL}$) give results close of 9.8, 2.7 and 3.4 [9.9, 2.7 and 3.3] Hz respectively for $J(C1-C2)$, $J(C2-C3)$ and $J(C3-C4)$ for the DTA [LTA] isomer. The $J(C-C)$ results for MTA isomer show less variation when we compare gas phase and solution results. For this isomer $\Delta J_{POL}(C1-C2)$, $\Delta J_{POL}(C2-C3)$ and $\Delta J_{POL}(C3-C4)$ values are around -2.1, 2.2 and -0.7 Hz. For the latter case, we observed different behavior for the values of the terminal carbons coupling constants. Only for these we have a reduction in the $J(C-C)$ value with the inclusion of the solvent. Results with PCM model are qualitatively/quantitatively the same observed with POL model.

### 3.5.1 Solvent effects on $J(C-O)$.

Table 3 show the $J(C-O)$ values for gas phase and in-water solution. For $J(C-O)$ of the isomers DTA and LTA we obtained similar results again. The $\Delta J_{POL}$ results are close of -4.6, 3.4, -1.9, 0.3, -2.2 and 1.3 [-4.1, 3.4, -1.9, 0.4, -2.5 and 1.1] Hz, respectively for $\Delta J_{POL}(C1-O8)$, $\Delta J_{POL}(C1-O9)$, $\Delta J_{POL}(C2-O13)$, $\Delta J_{POL}(C3-O15)$, $\Delta J_{POL}(C4-O7)$ and $\Delta J_{POL}(C4-O11)$ for the DTA [LTA] isomer. These results show that there is a fluctuation in the behavior of the $J(C-O)$ values with inclusion of solvent medium. We observe that the gas phase values of the $J(C1-O9)$, $J(C3-O15)$ and $J(C4-O11)$ are increased in the solvent medium and $J(C1-O8)$, $J(C2-O13)$ and $J(C4-O7)$ are decreased. The MTA isomer has a different behavior, all $J(C-O)$ values are lower in solution than in the gas phase. For this case, the results for $\Delta J_{POL}$ are equal to -1.9, -0.6, -2.1, -1.0, -2.6 and -1.5 Hz, respectively. This differences in the behavior of $J(C-O)$ values of the TA isomers show that DTA and LTA are magnetically different to the MTA. We also emphasize that although small, these $\Delta J_{POL}$ values are easily noticeable experimentally. We also note that PCM results do not take the same findings observed with POL model for several $J(C-O)$ analyzed.

## 4. CONCLUSION

In this paper we evaluate the behavior of magnetic shielding and spinspin coupling constants for carbon and oxygen atoms of the isomers of tartaric acid. We considered the influence of the polarization effect of tartaric acid in water solution using the sequential Monte Carlo / Quantum Mechanics methods. Our results demonstrate that the dipole moment of isomers are increased on average between 61-221% when compared to gas-phase results. In this regard, this polarization effects provide reductions in gas phase $\sigma(C)$ values for carbon atoms. This solvent shift is relevant and can reach up to 13 ppm. For oxygen atoms we observe a greater influence of the solvent effects on the values of gas phase $\sigma(O)$. In this case, we observe a solvent shift of 65 ppm when we use super molecular models, which explicitly solvent molecules in the quantum mechanics calculations. We also evaluated the spinspin coupling constant behavior between carbon-carbon atoms and between carbon-oxygen atoms. This property is more sensitive than the magnetic shielding constants and our results show that the differences between solution and gas phase results ($\Delta J_{POL}$) give results close of 9.8, 2.7 and 3.4 [9.9, 2.7 and 3.3] Hz respectively for $J(C1-C2)$, $J(C2-C3)$ and $J(C3-C4)$ for the DTA [LTA] isomer. These results emphasize the importance of inclusion of solvent effects to obtain realistic results of electrical and magnetic properties.